\documentstyle[amssymb,prb,aps,psfig]{revtex}

\begin{document}
\title{Glassy dynamics of the inhomogeneous metallic phase in La$_{1-x}$Ca$_{x}$MnO$%
_{3}$}
\author{F. Cordero,$^{1}$ C. Castellano,$^{2}$ R. Cantelli,$^{2}$ and M. Ferretti$%
^{3}$}
\address{$^{1}$ CNR, Istituto di Acustica ``O.M. Corbino``, Via del Fosso del
Cavaliere 100,\\
I-00133 Roma, Italy and Unit\`{a} INFM-Roma1, P.le A. Moro 2,I-00185 Roma,
Italy}
\address{$^{2}$ Universit\`{a} di Roma ``La Sapienza``, Dipartimento di Fisica, and\\
Unit\`{a} INFM-Roma 1, P.le A. Moro 2, I-00185 Roma, Italy}
\address{$^{3}$ Universit\`{a} di Genova, Dipartimento di Chimica e Chimica Fisica,\\
Via Dodecanneso 31, I-16146 Genova, and INFM, Italy}
\maketitle

\begin{abstract}
The anelastic spectra (elastic energy loss and dynamic modulus versus
temperature) of La$_{1-x}$Ca$_{x}$MnO$_{3}$ have been measured for $0.33<$ $%
x<$ $0.5$. A peak in the imaginary part of the elastic susceptibility is
found slightly below the Curie temperature, whose temperature and frequency
dependences are typical of the dielectric and magnetic susceptibilities of
relaxor ferroelectrics, spin glasses and other inhomogeneous frustrated
systems. The fluctuations can be associated with the magnetic clusters which
are observed by several techniques or with charge-ordered insulating domains
in the low temperature inhomogeneous phase. The temperature dependent
distribution function of the fluctuation energy barriers has been extracted
by the analysis of the experimental curves.
\end{abstract}

\twocolumn

The perovskite manganese oxides R$_{1-x}$A$_{x}$MnO$_{3}$ (R = La, Pr, or Nd
and A = Ca, Ba, Sr, or Pb) are the object of intense study, since the strong
coupling between their electronic, magnetic, orbital and structural degrees
of freedom gives rise to a rich phase diagram as a function of doping and
temperature.\cite{SRB95} In particular, La$_{1-x}$Ca$_{x}$MnO$_{3}$ in the
doping range $0.20<$ $x<$ 0.50 undergoes a phase transition from
paramagnetic insulating (PI) to metallic ferromagnetic (FM) upon cooling
below room temperature, and exhibit the so called negative colossal
magnetoresistance (CMR). The original interpretation of the
magneto-conductive phase transition by the classical double exchange (DE)
mechanism has been recently demonstrated to be not quantitatively
sufficient. Therefore a strong charge-phonon polaronic coupling, associated
with the Jahn-Teller distortion of the Mn$^{3+}$(O$^{2-}$)$_{6}$\ octahedra,
has been supposed\cite{MLS95}, which still does not supply an exhaustive
interpretation of CMR. In particular, a strong magnetic correlation between
charge carriers and local moments causes the self-trapping of the carriers
in magnetic polarons\cite{Nag00} over regions of about 10-20 \AA . When a
magnetic field is applied, these ferromagnetic clusters overlap leading to
carrier delocalization and consequently to CMR.\cite{Nag00,MCK99} Such spin
polarons have been first revealed in non Jahn-Teller systems presenting CMR,
like dilute magnetic semiconductors, rare earth chalcogenides, Mn
pyrochlores, and now also in manganites around the Curie temperature $T_{%
\text{C}}$ by inelastic neutron scattering,\cite{LEB96} small angle neutron
scattering\cite{TIA97} (SANS) and NMR.\cite{ADL98} These magnetic polarons
can be the precursors of the phase inhomogeneities recently evidenced below $%
T_{\text{C}}$ by STM,\cite{FFM99} NMR\cite{PFB00} and X-ray diffraction
studies.\cite{BPP00} In this phase, insulating and metallic domains of 10-50
nm coexist and percolate as a function of temperature and applied magnetic
field. Such a phase separated scenario recalls the relaxor ferroelectrics,
which also are perovskites, where the mixed composition induces
ferroelectric clusters of nanometer dimensions embedded in a paraelectric
matrix.\cite{VJC90,LC95} Analogously, in manganites with mixed compositions
lile La$_{1-x-y}$Pr$_{y}$Ca$_{x}$MnO$_{3}$ and Nd$_{0.5}$Ca$_{0.5}$Mn$_{1-y}$%
Cr$_{y}$O$_{3}$ it is possible to induce the coexistence of domains of FM\
and charge-orbital ordered (CO-OO) phases, which are stable and visible by
electron diffraction and by electron and magnetic force microscopy.\cite
{KTK99,UMC99} In these cases aging effects in magnetization and conductance
have been observed just like those revealed in glassy magnetic or relaxor
ferroelectrics. It seems that in the CMR manganites this phase separation is
more overbalanced towards the ferromagnetic component and the insulating CO
component is more fluctuating and less coherent, resulting therefore not
visible by diffraction. Increasing the Ca doping above 0.50, the CO phase
becomes long ranged.\cite{SRB95}

In this general context it is of great interest to study the dynamics of
these inhomogeneous fluctuating domains. This study is possible by anelastic
spectroscopy measurements, since the elastic constants and the elastic
energy dissipation coefficient are strictly connected to the dynamic
response of the system,\cite{NB} and also the magnetic fluctuations can
affect the anelastic response through magnetoelastic coupling.

The La$_{1-x}$Ca$_{x}$MnO$_{3}$ samples, with $x$ ranging from 0.33 to 0.5,
were prepared from La$_{2}$O$_{3}$, CaCO$_{3}$ and MnO$_{2}$ powders and
sintered at 1300-1400~$^{\text{o}}$C several times, until the X-ray
diffraction spectra indicated that all the starting powders had reacted; the
ingots were cut into bars approximately $45\times 4\times 0.5$~mm$^{3}$. The
complex dynamic Young's modulus $E\left( \omega ,T\right) =E^{\prime
}+iE^{\prime \prime }$ was measured as a function of temperature by
electrostatically exciting the 1st and 5th free flexural modes, whose
frequencies $\omega /2\pi $ are in the ratios $1:13.2$ and detecting the
vibration amplitude with a frequency modulation technique. The relative
change of the real part $E^{\prime }$ was measured from the temperature
dependence of $\omega ^{2}=$ $E^{\prime }/\rho $, where the mass density $%
\rho $ varies much less than $E$; the imaginary part is related to the
elastic energy loss coefficient (or reciprocal of the mechanical $Q$) by $%
Q^{-1}\left( \omega ,T\right) =E^{\prime \prime }/E^{\prime }=$ $\chi
^{\prime \prime }/\chi ^{\prime }$, where the $\chi =1/E$ is the elastic
susceptibility (compliance), and it was measured from the decay of the free
oscillations or from the width of the resonance peak.

The anelastic spectrum of La$_{1-x}$Ca$_{x}$MnO$_{3}$ with $0.33\le x\le 0.5$
is dominated by the trigonal/orthorhombic structural transition at $T_{\text{%
t}}$ and the PI/FM transition at $T_{\text{C}}$. Figure 1 presents the whole
anelastic spectrum of a sample with $x=0.33$, with $T_{\text{t}}=721$~K and $%
T_{\text{C}}=250$~K. The sharpness of the structural transition at $T_{\text{%
t}}=721$~K is an indication of the good sample homogeneity, while the huge
stiffening below $T_{\text{C}}=250$~K demonstrates the strong interplay
between orbital/magnetic and lattice phenomena, already observed by
ultrasonic measurements.\cite{RSC96,ZZS99} Similar spectra were obtained
with Ca doping up to 50\%.

In the present paper we are concerned with the region of the spectrum just
below the PI/FM transition, which did't present any particular feature in
the previous ultrasonic experiments.\cite{RSC96,ZZS99} This part of the
spectrum is shown in Fig. 2 for the sample with $x=0.33$. In analyzing
acoustic data, it is always assumed that the absorption can be identified
with the imaginary susceptibility; in the present case, however, we have to
analyze the spectrum just below $T_{\text{C}}$, where the large and rapid
stiffening of the modulus has to be taken into account. Therefore, instead
of $Q^{-1}$ we plot $\chi ^{\prime \prime }/\chi _{0}=Q^{-1}\chi ^{\prime
}/\chi _{0}$, where $\chi ^{\prime }\left( \omega ,T\right) =$ $\chi
_{0}\left[ \omega \left( 0\right) /\omega \left( T\right) \right] ^{2}$ and $%
\chi _{0}=\chi ^{\prime }\left( 0,0\right) $ is the reciprocal of the static
Young's modulus at $T = 0$. Beside a sharp attenuation peak centred near $T_{%
\text{C}}=250$~K, also determined from the onset of the stiffening (Fig. 1)\
and from the peak in the electrical resistivity, a broad shoulder appears at
lower temperature. Such a peak is the focus of the following discussion, and
was observed, although with different amplitudes, in all our samples with
five Ca doping levels included between 0.33 and 0.5; we chose to analyze the
two cases in which the background subtraction was easier. First of all the
absorption connected with the PI/FM transition has to be subtracted. The
peak cannot be identified with the critical absorption,\cite{NB,ZZS99} $%
Q_{c}^{-1}\propto \omega \left( T-T_{\text{c}}\right) ^{-\theta }$, which
was found\cite{ZZS99} in La$_{0.67}$Ca$_{0.33}$MnO$_{3}$ at 10~MHz with $%
\theta =1.55$; indeed, the amplitude of that peak extrapolated to our
frequencies becomes smaller than $1\times 10^{-5}$, while the height of the
peak in Fig. 2 is $5-8\times 10^{-4}$ (but is smaller in the samples with
higher Ca doping). Whatever the precise origin of the absorpion around $T_{%
\text{C}}$, it can be well fitted with a Lorentzian centred near $T_{\text{C}%
}$, as shown by the continuous lines in Fig. 2a. After subtraction of the
absorption around $T_{\text{C}}$ and of a small linear background, the two
curves in Fig. 2b are obtained. The resulting peak shifts to higher
temperature with increasing frequency, like a Debye relaxation process with
a thermally activated relaxation time $\tau $, usually following the
Arrhenius law, $\tau =\tau _{0}\exp \left( E/T\right) $. The response of
such a process would be\cite{NB} 
\begin{equation}
\chi \left( \omega ,T\right) \propto T^{-1}\,\left( 1-i\omega \tau \right)
^{-1}\,,  \label{Debye}
\end{equation}
whose imaginary part is peaked at a temperature $T_{m}$ satisfying the
condition 
\begin{equation}
\omega \tau \left( T_{m}\right) =1\,.  \label{wt=1}
\end{equation}
However, in that case the whole dispersion would shift to higher temperature
with increasing frequency, whereas the low temperature side of the peak in
Fig. 2b, is independent of frequency. In addition, the height of the peak
increases considerably with temperature, instead of decreasing as $1/T$. All
these features are typical of the dispersion of the imaginary susceptibility
of spin-glasses, mixed ferroelectric/antiferroelectric systems\cite{Cou84}
or relaxor ferroelectrics,\cite{VJC90,LC95} and therefore the absorption
peak seems to originate from the glassy dynamics of some inhomogeneous
phases, for example from the formation and disappearance or reorientiation
of the minority insulating nanodomains which are believed to coexist with
the FM phase below $T_{\text{C}}$. In order to extract a more quantititative
signature of the glassy dynamics, we performed on the present data the type
of analysis adopted for spin-glasses and their analogues, including the
relaxor ferroelectrics.

At first we tested whether the data at different frequencies can be
collapsed on a single curve by means of an appropriate scaling variable $%
E\left( \omega ,T\right) $ related to the law of the slowing down of the
fluctuations. The idea is that the susceptibility can be expressed in terms
of superposition of elementary processes, with a very broad distribution $%
g\left( \tau ,T\right) $ of relaxation times $\tau $, 
\begin{equation}
\chi \left( \omega ,T\right) =\chi \left( 0,T\right) \int_{\tau
_{0}}^{\infty }d\left( \ln \tau \right) \,g\left( \tau ,T\right) \frac{1}{%
1-i\omega \tau }  \label{superpos}
\end{equation}
so that each contributing peak in the imaginary part can be approximated as
a $\delta $ function, $\omega \tau /\left[ 1+\left( \omega \tau \right)
^{2}\right] \rightarrow $ $\frac{\pi }{2}\delta \left( \omega \tau -1\right)
.$ Then $\chi ^{\prime \prime }\left( \omega ,T\right) \simeq \frac{\pi }{2}%
\chi \left( 0,T\right) g\left( \omega ^{-1},T\right) $ is directly related
to $g$ through a scaling relationship often derived from the Vogel-Fulcher
law 
\begin{equation}
\tau =\tau _{0}\exp \left[ E/\left( T-T_{0}\right) \right] \,  \label{VF}
\end{equation}
and the condition (\ref{wt=1}). Following Courtens,\cite{Cou84} we tested
the data against the expression 
\begin{equation}
\chi ^{\prime \prime }=f\left( T\right) R\left( E_{c}-E\right)  \label{Cou}
\end{equation}
where $R\left( x\right) =$ $\left[ 1+\tanh \left( bx\right) \right] /2$
produces the high temperature dispersion, $E_{c}$ is a constant and $E$ is
related to $\omega $ and $T$ through Eqs. (\ref{wt=1}) and (\ref{VF}). The
function $f\left( T\right) $ is determined by fitting the frequency
independent region of $\chi ^{\prime \prime }$, and is found to be well
approximated by $f\left( T\right) =A\exp \left( W/T\right) $ with $W=4460$%
~K. The dark continuous curves of Fig. 2b, obtained with $\tau
_{0}=5.5\times 10^{-12}$~s, $T_{0}=193$~K, $E_{c}=620$~K and $b=0.011$,
demonstrate the glassy nature of the relaxation process just below $T_{\text{%
C}}$, and the parameter $T_{0}$ could be interpreted as a freezing
temperature; the use of a simple Arrhenius law, putting $T_{0}=0$, yielded a
definitely worse fit.

The same analysis has been carried out for a sample with $x=0.50$ nominally,
but with actual Ca doping slightly below that value, since only the PI/FM
transition was detected, and the resistivity curve did not show the expected
insulating behavior at low temperature. At this doping the intensity of the
glassy relaxation peak was higher, but the background subtraction was less
obvious, due to the presence of another peak at lower temperature; for this
reason, the simultaneous optimization of the background and peak parameters
was preferred. Figure 3 shows the resulting fit with the same criteria used
for the $x=0.33$ case. The resulting parameters are $W=1920$~K, $\tau
_{0}=6.4\times 10^{-8}$~s, $T_{0}=180$~K, $E_{c}=272$~K and $b=0.0124$.

While the above fits demonstrate the close similarity between the dynamics
of the inhomogeneities in La$_{1-x}$Ca$_{x}$MnO$_{3}$\ below $T_{\text{C}}$\
and a freezing process, most of the parameters have little physical meaning.
In order to obtain a more physical picture, we also carried out another type
of analysis, which was recently adopted to describe the dielectric
susceptibility of relaxor ferroelectrics.\cite{LC95} The relaxation process
is described in terms of a distribution of activation energies $g\left(
E,T\right) =$ $E_{0}^{-1}\exp \left[ \left( E_{c}-E\right) /E_{0}\right] $
with lower cutoff at $E_{c}$ and a temperature dependent width $E_{0}\left(
T\right) $. The integration of the elementary relaxation function Eq. (\ref
{Debye}) over $g\left( E,T\right) $, assuming the Arrhenius law $\tau =$ $%
\tau _{0}\exp \left( E/T\right) $, yields a very simple expression\cite{LC95}
of $\chi ^{\prime \prime }$, which we generalized including the
Vogel-Fulcher law, Eq. (\ref{VF}). The result is 
\begin{eqnarray}
\chi ^{\prime \prime } &\simeq &\chi _{0}\,\frac{\pi n}{2\cos \left( n\pi
/2\right) }\exp \left( E_{c}/E_{0}\right) \left( \omega \tau _{0}\right) ^{n}
\label{LC-VF} \\
\text{with }n &=&\frac{T-T_{0}}{E_{0}}\,.
\end{eqnarray}
The simple result (\ref{LC-VF}) is valid for $n\le 1$, and therefore we put
an upper cutoff to $n$, which however does not affect the fit for the case $%
x=0.33$, since $n$ approaches 1 only in the high temperature region where
the curves are dropping to zero. The temperature dependence of $E_{0}$ can
be extracted from the frequency independent part of the curves at low
temperature, where $n\ll 1$ and, initially assuming $T_{0}=0$ for
simplicity, $\chi ^{\prime \prime }\left( T\right) \simeq \left( \pi
/2\right) \,\chi _{0}T/E_{0}$; as already mentioned, in that region $\chi
^{\prime \prime }\left( T\right) \simeq $ $A\exp \left( W/T\right) $, and
therefore we adopted $E_{0}\left( T\right) =$ $cT\exp \left( W/T\right) $,
where $c$ is a constant. Since $W=2000-4500$~K, we are in the regime were
the width $E_{0}\left( T\right) $ of the distribution function diverges with
lowering temperature, as required by a freezing process. The gray curves of
Fig. 2b are obtained with \ $W=4480$~K, $c=2.5\times 10^{-9}$, $E_{c}=880~$%
K, $\tau_{0}=4\times 10^{-14}$~s and $T_{0}=186$~K. The fit is satisfactory,
considering the simplificative hypotheses made for the distribution function
of the activation energies, $g\left( E,E_{0}\right) $ with $E_{0}\left(
T\right) $, and characterizes the spectrum of the fluctuations. The
relaxation rates are close to $\tau _{0}^{-1}\exp \left( -E_{c}/T\right) $
at high temperature, but the spectrum quickly broadens with decreasing
temperature, consistently with a freezing temperature $T_{0} \simeq 186$~K.
In a sense, the broadening of $g\left( E,T\right) $ itself should describe
the freezing process, without the need of introducing the Vogel-Fulcher law,
but the limitation of using a tractable expression for $g\left( E,T\right) $
makes it useful the introduction of $T_{0}>0$ for improving the fit. For $%
x=0.5$ (Fig. 3) the condition $n\le 1$ for the validity of Eq. (\ref{LC-VF})
was shifted to slightly lower temperature, making the fit less meaningful
for the data at higher frequency. Anyway, the case $x=0.5$ can be described
by $W\gtrsim $ 1900~K, $E_{c}\gtrsim $ $980$~K, $\tau _{0}\le 2\times
10^{-10}$~s; the use of $T_{0}>0$ does not improve significantly the fit.

Let us consider now the nature of the fluctuations which are observed in the
anelastic spectra of La$_{1-x}$Ca$_{x}$MnO$_{3}$. A\ first possibility are
the FM\ clusters which are observed on different scales, with the anelastic
response driven by strong magnetoelastic coupling. Indeed, around $T_{\text{C%
}}$ SANS measurements\cite{TIA97} reveal the presence of ''magnetic
polarons'' with sizes of the order of tens of \AA , while in neutron
spectroscopy experiments at $x=0.33$ a quasielastic components appears,
which indicate a diffusive component of the spin excitations, coexisting
with the usual spin waves of a typical ferromagnet.\cite{LEB96} It should be
noted that, although the present measurements probe the very low frequency
part of the fluctuation spectrum, the resulting distribution of activation
energies $g\left( E,T\right) $ acquires weight near $E_{c}\simeq 880$~K with
increasing temperature above 220~K; this corresponds to very fast relaxation
rates, in the range of $10^{-12}$~s$^{-1}$, which are the rates probed by
neutron spectroscopy. In addition, the relaxation could be connected with
the fluctuation of the magnetization in the nanometric FM domains, observed
by STM\cite{FFM99} and suggested by PDF XRD\cite{BPP00} and NMR\cite{PFB00}
experiments. Another possibility is that we observe the CO insulating
clusters which coexist with the FM\ phase, and could consist of diagonal Mn$%
^{4+}$/Mn$^{3+}$ stripes, as observed\cite{RCM97} in the CO coherent phase
at $x\ge 0.50$. In this case the anelastic response may result both from the
formation and disappearance of the CO state over few lattice constants or
from the reorientation of the Jahn-Teller distorted stripes.

Regarding the origin of the glassy dynamics, in some cases a broad
distribution of the activation energies is supposed to reflect a broad
distribution of the sizes of the clusters.\cite{LC95} We rather suggest that
the main cause of the glassy dynamics is the competition of interactions
with different length scales, which may produce spatial inhomogeneities with
well defined patterns but with glassy dynamics.\cite{SW00} In fact, the
relevant interactions in the manganites range from the magnetic exchange to
the Coulomb and elastic ones. The broadening of the distribution function $%
g\left( \tau ,T\right) $ with lowering $T$ can then be due to the frustrated
nature of the interactions between the coexisting insulating and metallic\
regions.

In conclusion, the low frequency dynamic susceptibility of La$_{1-x}$Ca$_{x}$%
MnO$_{3}$, probed by anelastic spectroscopy for $0.33<$ $x<$ $0.5$, contains
the signature of glassy dynamics in the low temperature inhomogeneous
ferromagnetic phase. A peak in the elastic energy loss is found slightly
below the Curie temperature, whose temperature and frequency dependences are
typical of the dynamic susceptibilities of other perovskite inhomogeneous
frustrated systems like the relaxor ferroelectrics, and of spin glasses. The
data have been analyzed in order to extract the temperature dependent
distribution function of the energy barriers of the fluctuations
characterizing the inhomogeneous state.

Captions to figures

\begin{figure}[]
%\vspace{4.5 cm}
\caption{Elastic energy loss coefficient, $E^{\prime\prime}/E^{\prime}$, and
normalized real part of the dynamic Young's modulus of La$_{0.67}$Ca$_{0.33}$%
MnO$_{3}$.}
\label{fig1}
\end{figure}

\begin{figure}[]
%\vspace{12 cm}
\caption{Imaginary part of the elastic susceptibility of La$_{0.67}$Ca$%
_{0.33}$MnO$_{3}$ measured at two vibration frequencies. The continuous
lines in (a) have been subtracted to obtain the peak in (b). The black
curves in (b) are the fit according to Eq. (\ref{Cou}), while the gray ones
according to Eq. (\ref{LC-VF}).}
\label{fig2}
\end{figure}

\begin{figure}[tbp]
%\vspace{6.3 cm}
\caption{ Imaginary part of the elastic susceptibility of La$_{1-x}$Ca$_{x}$%
MnO$_{3}$ with $x$ slightly below 0.50. The fit is made as described in the
text, according to Eq. (\ref{Cou}).}
\label{fig3}
\end{figure}

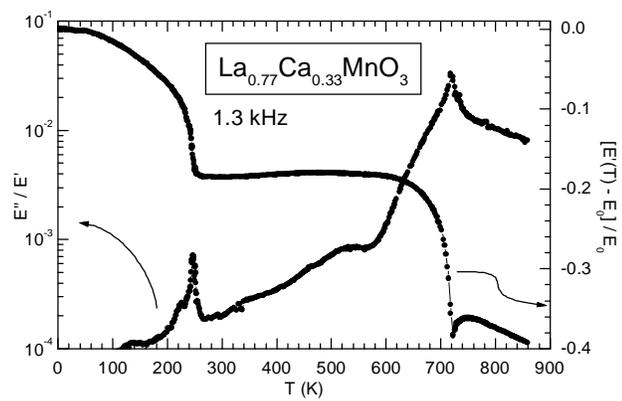
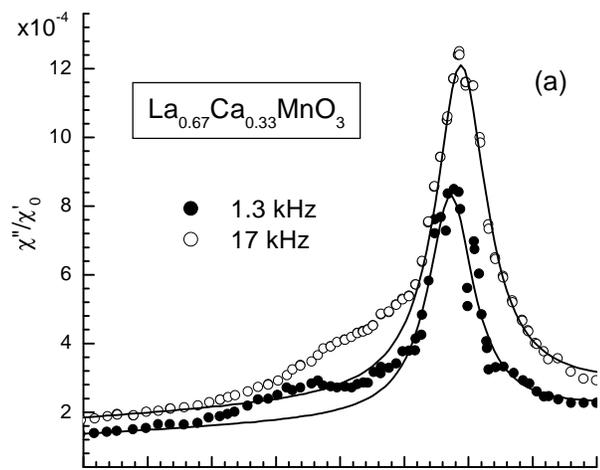
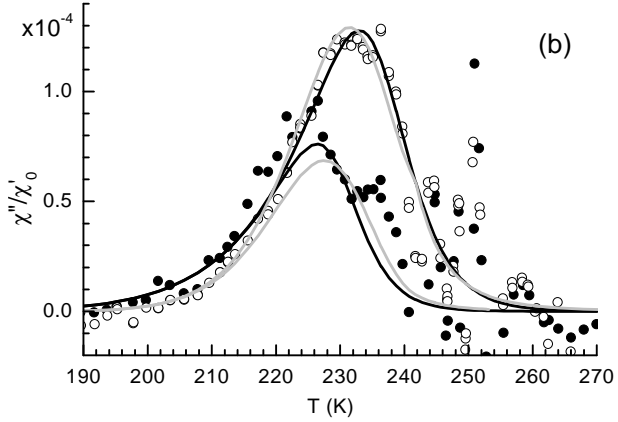
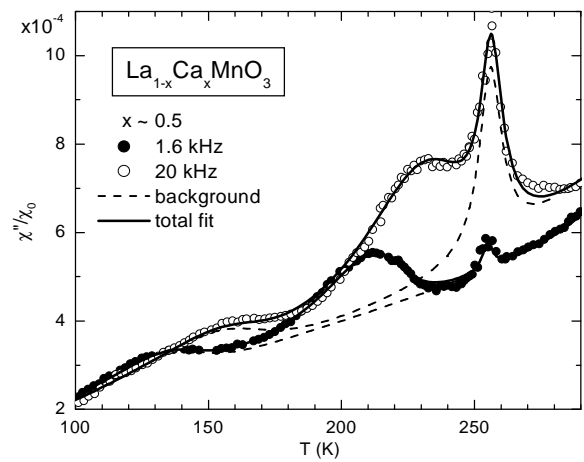
\end{document}